\documentclass[english]{article}
\usepackage[T1]{fontenc}
\usepackage[latin9]{inputenc}
\usepackage[a4paper]{geometry}
\usepackage{babel}
\usepackage{units}
\usepackage{amstext}
\usepackage{graphicx}
\usepackage[authoryear]{natbib}
\usepackage[unicode=true,
 bookmarks=true,bookmarksnumbered=false,bookmarksopen=false,
 breaklinks=false,pdfborder={0 0 1},backref=false,colorlinks=false]
 {hyperref}
\hypersetup{pdftitle={Functional responses in resource-based mutualisms: a time scale approach},
 pdfauthor={Tomas Revilla},
 pdfkeywords={mutualism, resources, services, steady-state, functional response}}
\usepackage{breakurl}

\makeatletter
\usepackage[auth-lg]{authblk}
   \author[1]{Tomás A. Revilla}
   \affil[1]{Centre for Biodiversity Theory and Modelling, 09200 Moulis, France}

\usepackage{lineno}

\makeatother

\begin{document}

\title{Numerical responses in resource-based mutualisms: a time scale approach}
\maketitle
\begin{abstract}
In mutualisms where there is exchange of resources for resources,
or resources for services, the resources are typically short lived
compared with the lives of the organisms that produce and make use
of them. This fact allows a separation of time scales, by which the
numerical response of one species with respect to the abundance of
another can be derived mechanistically. These responses can account
for intra-specific competition, due to the partition of the resources
provided by mutualists, in this way connecting competition theory
and mutualism at a microscopic level. It is also possible to derive
saturating responses in the case of species that provide resources
but expect a service in return (e.g. pollination, seed dispersal)
instead of food or nutrients. In both situations, competition and
saturation have the same underlying cause, which is that the generation
of resources occur at a finite velocity per individual of the providing
species, but their depletion happens much faster due to the acceleration
in growth rates that characterizes mutualism. The resulting models
can display all the basic features seen in many models of facultative
and obligate mutualisms, and they can be generalized from species
pairs to larger communities. The parameters of the numerical responses
can be related with quantities that can be in principle measured,
and that can be related by trade-offs, which can be useful for studying
the evolution of mutualisms.

\end{abstract}

\section*{Introduction}
\begin{quote}
\begin{flushright}
\emph{Nous ne notons pas les fleurs, dit le géographe}
\par\end{flushright}

\begin{flushright}
\emph{Pourquoi ça! c'est le plus joli!}
\par\end{flushright}

\begin{flushright}
\emph{Parce que les fleurs sont éphémères}
\par\end{flushright}

\begin{flushright}
Le Petit Prince, Chapitre XV -- \citet{petitprince}
\par\end{flushright}
\end{quote}
Early attempts to model the dynamics of mutualisms were based on phenomenological
descriptions of interactions. The best known example involves changing
the signs of the competition coefficients of the Lotka-Volterra model,
to reflect the positive contribution of mutualism \citep{vandermeer_boucher-jtb78,may1981}.
This simple, yet insightful approach, predicts several outcomes depending
on whether mutualism is facultative or obligatory. One example is
the existence of population thresholds, where populations above thresholds
will be viable in the long term, but populations below will go extinct.
The same approach however, reveals an important limitation, that the
mutualists can help each other to grow without limits, in an ``orgy
of mutual benefaction'' (sic. \citealp{may1981}), yet this is never
observed in nature. One way to counter this paradox is to assume that
mutualistic benefits have diminishing returns \citep{vandermeer_boucher-jtb78,may1981},
such that negative density dependence (e.g. competition) would catch
up and overcome positive density dependence (mutualism) at higher
densities. This makes intuitive sense because organisms have a finite
nature (e.g. a single mouth, finite membrane area, minimum handling
times, etc), causing saturation by excessive amounts of benefits.
Other approaches consider cost-benefit balances that change the sign
of inter-specific interactions from positive at low densities (facilitation)
to negative at high densities (antagonism) \citep{hernandez-rspb98}.

\citet{holland_deangelis-ecology10} introduced a general framework
to study the dynamics of mutualisms. In their scheme two species,
1 and 2, produce respectively two stocks of resources which are consumed
by species 2 and 1, according to Holling's type II functional response,
and which are converted into numerical responses by means of conversion
constants. In addition, they consider costs for the interaction in
one or both of the mutualists, which are functions of the resources
offered to the other species, also with diminishing returns. In their
analyses, the resources that mediate benefits and costs are replaced
by population abundances as if the species were the resources themselves.
This assumption enables the graphical prediction (nullcline analysis)
of a rich variety of outcomes, such as Allee effects, alternative
stable mutualisms, and transitions between mutualisms and parasitisms.

The work of \citet{holland_deangelis-ecology10} uses concepts consumer-resource
theory to study the interplay between mutualism and antagonism at
population and community levels, but the functional responses are
not derived from microscopic principles. In other words, there is
no explicit mechanism that explains why is that the resource provided
by species 1, can be replaced by the abundance of species 1 (or some
function of it). If the functional responses are considered phenomenologically
that is not a problem, consumer-resource theory makes predictions
using phenomenological relationships, like the Monod and Droop equations
\citep{grover1997}. For example, the \emph{half-saturation} for mutualism
in species 1 is a trivial concept, it is just the abundance of species
2 that reports one half of the maximum benefit that species 1 can
receive. But things can be conceptually problematic if one attempts
to explain other parameters. For example, what is the \emph{handling
time} of a plant that uses a pollinator or seed disperser? or at which
\emph{rate} does a plant \emph{attack} a service?

In this note I will show that in some scenarios of mutualism, it is
very convenient to consider the dynamics of the resources associated
with the interaction in a more explicit manner, before casting them
in terms of the abundances of the mutualists. As it turns out in many
situations, these resources have life times that are on average much
shorter than the lives of the individuals producing and profiting
from them. For example, the life of a tree can be measured in years
and that of a small frugivore in months, but many fruits do not last
more than a few weeks. This also apply for flowers in relation to
pollinators like hummingbirds, but certainly not to mayflies (so called
\emph{Ephemero}ptera because of their very short lives as adults).
Taking advantage of this fact, the resources can be assumed to attain
a steady-state, and thus be quantified in terms of the present abundances
of the providers and the consumers in a mechanistic manner. This approach
has been used extensively to derive some of the most important relationships
in biology, from enzyme kinetics \citep{briggs_haldane-biochemj25},
to the Lotka-Volterra model of competition itself \citep{macarthur-tpb70},
functional responses \citep{real-amnat77}, and the dynamics of infectious
diseases \citep{may_anderson-nature79}. Using this approach, it is
possible to derive not just the numerical responses in terms of populations
abundances, but also to do it in terms of parameters that could be
measured, such as the rates of resource production, their decay, and
consumption. Intra-specific competition for mutualistic benefits can
be related to consumption rates, and concepts such as the handling
time of a plant can be given a meaning. This in turn opens the possibility
of framing the costs of mutualism by means of trade-offs relating
vital parameters. The examples in this note are meant to promote more
thinking in this direction, that of considering separation of times
scales, in order to tie together mutualism, competition, and exploitation,
in more mechanistic ways.

\section*{Exchanges of resources for resources}

Consider two species $i,j=1,2$ that provide food to each other. Their
populations $(N_{i})$ change in time $(t)$ according to the differential
equations:

\begin{equation}
\frac{dN_{i}}{dt}=G_{i}(\cdot)N_{i}+\sigma_{i}\beta_{i}F_{j}N_{i}\label{eq:res-res}
\end{equation}

\noindent where $F_{j}$ is the amount of resources provided by species
$j$, $\beta_{i}$ is the per-capita consumption rate per unit resource
by species $i$, and $\sigma_{i}$ is the conversion ratio of eaten
resources into biomass by species $i$. The function $G_{i}$ is the
per-capita rate of change of species $i$ when it \emph{does not interact
with species $j$ by means of the mutualism}. The food dynamics is
accounted by a second set of differential equations:

\begin{equation}
\frac{dF_{j}}{dt}=\alpha_{j}N_{j}-\omega_{j}F_{j}-\beta_{i}F_{j}N_{i}\label{eq:food-food}
\end{equation}

Here I assume that the resource is produced in proportion to the biomass
of the provider with per-capita rate $\alpha_{i}$, and it is lost
or decays with a rate $\omega_{i}$ if it is not consumed. I also
assume that the physical act of eating by the receiver does not have
an negative impact such as damage or death, on the provider, e.g.
the food is secreted in the external environment. As stated in the
introduction, the lifetime of the food items are much shorter than
the dynamics of the populations, i.e. $\alpha_{i},\omega_{i}$ and
$\beta_{i}$ are large compared to $G_{i}$. This scenario allows
the food to achieve a steady-state or quasi-equilibrium dynamics before
the populations attain their long-term dynamics. Thus, assuming that
$dF_{j}/dt\approx0$ in equations (\ref{eq:food-food}), the steady-state
amount of food:

\begin{equation}
F_{j}\approx\frac{\alpha_{j}N_{j}}{\omega_{j}+\beta_{i}N_{i}}\label{eq:food}
\end{equation}

\noindent can be substituted in the dynamical equations of the populations
(\ref{eq:res-res}):

\begin{equation}
\frac{dN_{i}}{dt}=\left\{ G_{i}(\cdot)+\frac{\sigma_{i}\beta_{i}\alpha_{j}N_{j}}{\omega_{j}+\beta_{i}N_{i}}\right\} N_{i}\label{eq:bidir}
\end{equation}

In this model, the larger the receiver population, the lower the per-capita
rates of acquisition of mutualistic benefits. The decrease in returns
experienced by receiver $i$ happens because the resource produced
by the provider $(\alpha_{j}N_{j})$, must be shared among an increasing
numbers of individuals, each taking a fraction $\beta_{i}/(\omega_{j}+\beta_{i}N_{i})$.
This in effect describes intra-specific competition for a finite source
of energy or resources, as originally modeled by \citet{schoener-tpb78},
with the only difference that in Schoener's model the supply of resources
is at a constant rate.

\begin{figure}
\begin{centering}
\includegraphics{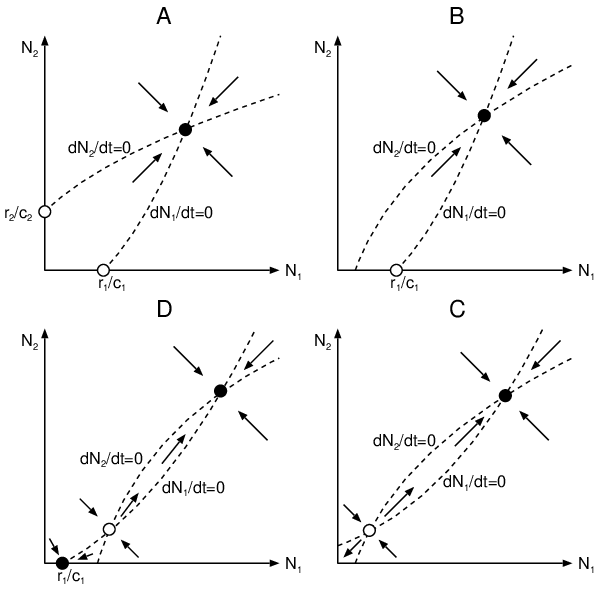}
\par\end{centering}

\caption{\label{fig:phaseplane1}Nullclines (dashed curves) and equilibrium
points in mutualisms with exchange of resources for resources, assuming
linear self-limitation for each species. Black and white circles represent
stable (nodes) and unstable (saddle) equilibria respectively (also
indicated by arrows nearby). A: When both species are facultative
mutualists, their nullclines always cross once giving rise to a single
globally stable mutualistic equilibrium. When species 1 is facultative
and species 2 is an obligate mutualist their nullclines may cross
as in B: once, giving rise to a single globally stable mutualistic
equilibrium; or as in C: twice, giving rise to an unstable and a locally
stable mutualistic equilibrium. When both species are obligate mutualists,
their nullclines may cross at two points (never a single one), an
unstable and a locally stable mutualistic equilibrium. The existence
of an unstable mutualism means that the obligate species (species
2 in C, both species in D) may go extinct depending on the initial
conditions or external perturbations. With the exception of case A,
the nullclines may also never cross, leading to the extinction of
one or both species (not shown).}

\end{figure}

Characterizing the system dynamics requires explicit formulations
of the growth rates in the absence of the mutualism, $G_{i}$. These
functions can range from very simple to very complicated depending
on the particular assumptions about the biology of species, the existence
of alternative food sources, whether mutualism is obligate or facultative,
the mechanisms of self-regulation, and the interactions (positive
or negative) with other species (even perhaps between species 1 and
2 by means other than mutualism). For illustration, I will consider
that $G_{i}$ is a linear decreasing function of species $i$ abundance
following \citet{holland_deangelis-ecology10}:

\begin{equation}
G_{i}(N_{i})=r_{i}-c_{i}N_{i}\label{eq:logistic}
\end{equation}

\noindent where $c_{i}>0$ is a coefficient of self-limitation and
$r_{i}$ is the intrinsic growth rate of $i$, which will be positive
for a facultative mutualist and negative for an obligate mutualist.
By substituting (\ref{eq:logistic}) in (\ref{eq:bidir}), turns out
that species $i$ increases $(dN_{i}/dt>0)$ only if:

\begin{equation}
N_{j}>\frac{(c_{i}N_{i}-r_{i})(\omega_{j}+\beta_{i}N_{i})}{\sigma_{i}\beta_{i}\alpha_{j}}\label{eq:nullcline}
\end{equation}

\noindent and decreases otherwise. With an equal sign, (\ref{eq:nullcline})
is the nullcline of species $i$ (zero-net-growth-isocline, formally).
The nullcline is an increasing parabola in the positive part of the
$N_{1}N_{2}$ plane. Species $i$ nullcline has two roots in the $N_{i}$
axis, one at $-\omega_{j}/\beta_{j}$ which is always negative, and
one at $r_{i}/c_{i}$ which is negative if $i$ is an obligate mutualist,
or positive if $i$ is a facultative mutualist. For a facultative
mutualist $r_{i}/c_{i}$ is also its carrying capacity, while for
an obligate mutualist $r_{i}$ could be its intrinsic mortality. Figure
\ref{fig:phaseplane1} shows the possible outcomes of the interaction,
which ranges from having a globally stable mutualistic equilibrium
when both species are facultative mutualists, to a locally stable
equilibrium and dependence on the initial conditions when one or both
species are obligate mutualists.

\begin{figure}
\begin{centering}
\includegraphics[clip,scale=0.6]{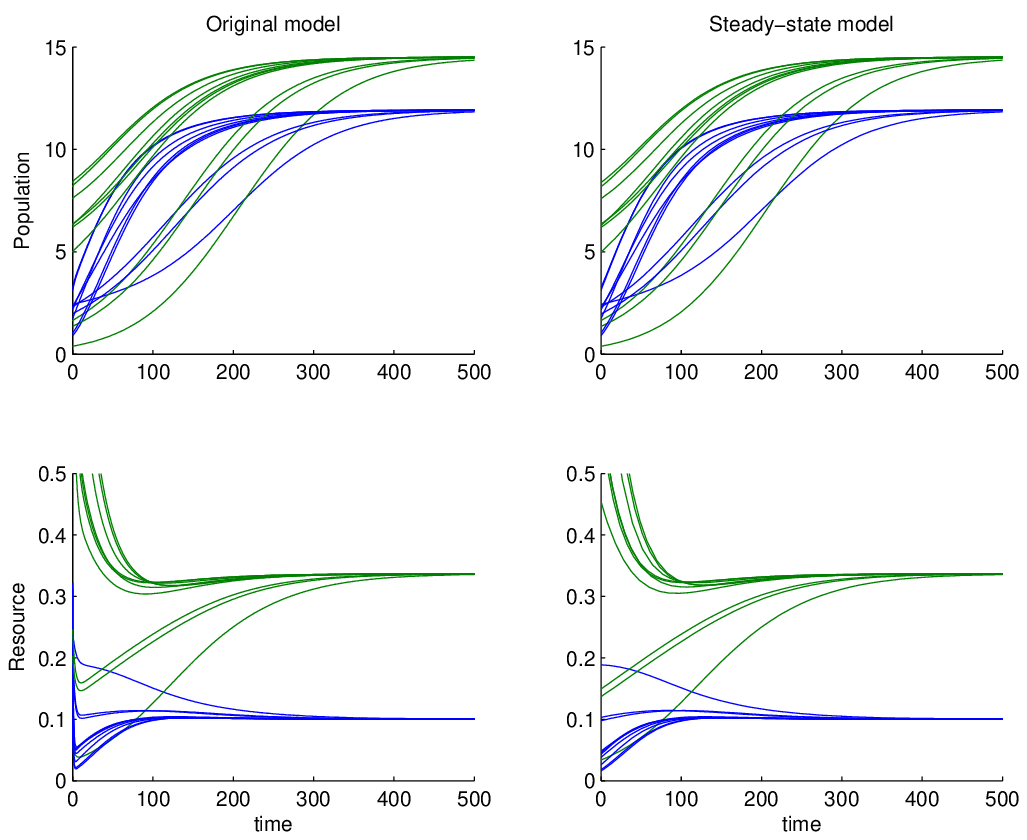}
\par\end{centering}

\caption{\label{fig:dynamics} Dynamics of the original mutualistic model (\ref{eq:res-res},\ref{eq:food-food})
in the left column, and of the steady-state model (\ref{eq:bidir})
in the right column. $G_{i}$ is defined by (\ref{eq:logistic}) in
both models. For each simulation in the original model a simulation
in the steady-state model is done using the same initial conditions.
Blue color is for species $i=1$ and green for $i=2$; $r_{i}=\{0.007,0.01\}$,
$c_{i}=\{0.002,0.001\}$, $\sigma_{i}=\{0.5,0.3\}$, $\alpha_{i}=\{0.02,0.03\}$,
$\omega_{i}=\{0.2,0.1\}$, $\beta_{i}=\{0.1,0.15\}$.}
\end{figure}

Figure \ref{fig:dynamics} show simulations of the original model
(\ref{eq:res-res}), in which the resource dynamics is explicitly
accounted by (\ref{eq:food-food}). The growth rates $r_{i}$ and
self-limitation coefficients $c_{i}$ were set at very low values
compared with flower production $\alpha_{i}$, decay $\omega_{i}$
and consumption $\beta_{i}$ rates, in some cases the differences
are more than two orders of magnitude. This makes resource dynamics
much faster than population dynamics. We can see that in the majority
of the cases, the amount of resources change very after a very short
time, compared with the time that it takes the populations to reach
an equilibrium. There a few instances of course, in which the resources
vary as much as the population densities during the same time frame.

Next to the simulation of the original model, the dynamics of the
model with resources under steady-state (\ref{eq:bidir}) is simulated
using the same parameters and initial conditions. For this model,
the resources follow equation (\ref{eq:food}) instead of a differential
equation. This leads to the same equilibria as the original model,
since the derivation of equation (\ref{eq:food}) is part of the procedure
to find the equilibrium in system (\ref{eq:res-res}). We also see
that the transient dynamics is very similar to the dynamics in the
original model. In fact, the differences between both models can be
chosen to be as little as desired, by widening the time scales between
population and resource dynamics.

Interestingly, using model (\ref{eq:bidir}), we can approximate the
dynamics of the resources. The derivation of (\ref{eq:food}) with
respect to time results in:

\begin{equation}
\frac{dF_{i}}{dt}=\alpha_{i}\frac{(\omega_{i}+\beta_{j}N_{j})\frac{dN_{i}}{dt}-\beta_{j}N_{i}\frac{dN_{j}}{dt}}{(\omega_{i}+\beta_{j}N_{j})^{2}}\label{eq:resapprox}
\end{equation}

Rather than substituting (\ref{eq:bidir}) above and complicating
it even more, it is more enlightening to consider the following two
scenarios. When both species have very low densities (\ref{eq:resapprox})
becomes $dF_{i}/dt\approx(\alpha_{i}/\omega_{i})dN_{i}/dt$, i.e.
the resources track the dynamics of the providers only. From Figure
(\ref{fig:dynamics}) we can see that this is a bad approximation,
because in many simulations the resources decrease despite the fact
that the populations are increasing. Now, if the populations are instead
large enough $(N_{j}\gg\omega_{i}/\beta_{j})$, equation (\ref{eq:resapprox})
is better approximated by:

\[
\frac{dF_{i}}{dt}\approx\left(\frac{\alpha_{i}}{\beta_{j}}\right)\frac{N_{j}\frac{dN_{i}}{dt}-N_{i}\frac{dN_{j}}{dt}}{N_{j}^{2}}
\]

At very large densities, population growth is forced to slow down
because of self-limitation. Indeed, before attaining an equilibrium,
populations ought to be growing at rates that are sub-exponential
(for an explanation see \citealp{szathmary-tree91}). We can see from
the simulations (Figure \ref{fig:dynamics}), that somewhere before
the decelerating phase, population growth rates can be approximated
as linear in time, i.e. $dN_{i}/dt\approx k_{i}$. We know that $dN_{i}/dt\approx0$
eventually, but let see what happens if we substitute linear growth
in the resource dynamics described in approximation above:

\[
\frac{dF_{i}}{dt}\approx\left(\frac{\alpha_{i}}{\beta_{j}}\right)\frac{k_{i}N_{j}-k_{j}N_{i}}{N_{j}^{2}}
\]

The numerator of this rate is linear and the denominator is quadratic.
It is easy to see that when the populations are large enough and growing
$dF_{i}/dt\to0$. In addition, since the populations will eventually
attain an equilibrium, $k_{i}$ and $k_{j}$ must also decrease in
time. Thus, the stabilization of the resources will happen even sooner
than the stabilization of the populations, and for this reason the
steady-state assumption is reasonably correct.

\section*{Exchanges of resources for services}

This time consider that only species 1 is the food provider, and species
2 gives a service to species 1 as a consequence of food consumption.
This situation occurs under pollination or in seed dispersal for example.
Thus, let us assume that species 1 is a plant and species 2 an animal.
The dynamical equations for plants and animals are:

\begin{eqnarray}
\frac{dN_{1}}{dt} & = & G_{1}(\cdot)N_{1}+\sigma_{o}\beta_{o}FN_{o}+\sigma_{1}\beta FN_{2}\nonumber \\
\frac{dN_{2}}{dt} & = & G_{2}(\cdot)N_{2}+\sigma_{2}\beta FN_{2}\label{eq:plant-animal}
\end{eqnarray}

In this scheme $F$ is the number of flowers or fruits produced by
the plant, and $\beta$ is the rate of pollination or frugivory by
the animal. The animal's equation is not different than before. The
plant's equation must be changed to reflect that plants do not eat
anything from the animals, they instead grow in proportion to the
number of flowers pollinated or fruits eaten (after all, they contain
ovules or seeds). As pollination or frugivory takes place at the rate
of $\beta FN_{2}$, the plants increase their chances of fertilization
or seed dispersal, with a yield $(\sigma_{1})$ that depends on the
balance of the many benefits and risks involved in their interaction
with the animals (e.g. seed mastication), as well as the number of
ovules per flower or seeds per fruit. The purpose of an additional
term $\sigma_{o}\beta_{o}FN_{o}$ is to account for the fact that
pollination and seed dispersal can happen in the absence of the animal
species explicitly considered, for example by the actions of another
animal (species ``\emph{o}'') or thanks to abiotic factors like
wind (then $\beta_{o},N_{o}$ would be proxies of e.g. wind velocity,
and $\sigma_{o}$ the corresponding yield).

Flower or fruit production is proportional to the plant's abundance,
and losses occur due to withering, rotting, pollination or consumption:

\begin{equation}
\frac{dF}{dt}=\alpha N_{1}-\omega F-\beta_{o}FN_{o}-\beta FN_{2}\label{eq:floresfrutas}
\end{equation}

The case of flowers deserves particular attention. Whereas a single
act of frugivory denies a fruit to other individuals \emph{ipso facto},
a single act of pollination will hardly destroy a flower. Certainly,
each pollination event brings a flower closer to fulfilling its purpose,
to close, and to stop giving away precious resources (nectar). Each
pollination event also makes a flower less attractive to other pollinators,
as it becomes less rewarding or damaged. This means that the decrease
in flower \emph{quantity} due to pollination $(\beta FN_{2})$ involves
a certain amount of decrease in \emph{quality}, rendering them useless
for plants and animals, a little bit each time. Thus, the pollination
rate in (\ref{eq:floresfrutas}) should be rather cast as $\kappa\beta FN_{2}$
where $0<\kappa\leq1$ is the probability that a flower stops working
as a consequence of pollination. This is a complication that is relevant
in specific scenarios, but it does not affect the generality of the
results derived, which is why it is not considered (so $\kappa=1$).
Similar to the previous scenario, assume that acts of pollination
or frugivory do not entail damage for individual plants, notwithstanding
the fact that flowers and fruits are physically attached to them.

Like before, consider that flowers or fruits are ephemeral compared
with the lives of plants and animals. Thus $F$ will rapidly attain
a steady-state $(dF/dt\approx0)$ compared with the much slower demographies.
The number of flowers or fruits can be cast a function of plant and
animal abundances $F\approx\alpha N_{1}/(\omega+\beta_{o}N_{o}+\beta N_{2})$,
and the dynamical system (\ref{eq:plant-animal}) as:

\begin{eqnarray}
\frac{dN_{1}}{dt} & = & \left\{ G_{1}(\cdot)+\frac{\sigma_{o}\beta_{o}\alpha N_{o}+\sigma_{1}\beta\alpha N_{2}}{\omega+\beta_{o}N_{o}+\beta N_{2}}\right\} N_{1}\nonumber \\
\frac{dN_{2}}{dt} & = & \left\{ G_{2}(\cdot)+\frac{\sigma_{2}\beta\alpha N_{1}}{\omega+\beta_{o}N_{o}+\beta N_{2}}\right\} N_{2}\label{eq:unidir}
\end{eqnarray}

\noindent where not surprisingly, the equation for the animal is practically
the same as in the previous model where both species provide resources
to each other. The only novelty is that the animal is sharing plant
resources with other processes or animals (quantified by $\beta_{o}N_{o}$),
thus hinting at competitive effects (more about this in the discussion).
The equation for the plant is however very different, because it includes
a saturating numerical response with respect to the abundance of its
mutualistic partner, and it is in fact a multispecific numerical response
if $N_{0}$ is the abundance of a different animal (more about this
in the discussion, again).

Let us assume for the moment that the plant relies exclusively on
species 2 for pollination or dispersal services (or $N_{o}=0$). Dividing
numerator and denominator by $\beta$, the numerical response of the
plant can be written in the Michaelis-Menten form:

\begin{equation}
\frac{vN_{2}}{K+N_{2}}=\frac{\alpha N_{2}}{\left(\nicefrac{\omega}{\beta}\right)+N_{2}}\label{eq:michaelismenten}
\end{equation}

\noindent where the maximum rate at which a plant acquires benefits
$v=\alpha$ is limited by the rate at which it can produce fruits
or flowers, and the half-saturation constant $K=\omega/\beta$ is
the ratio of the rate at which flower or fruits are wasted rather
than used by the animal, in other words a quantifier of \emph{inefficiency}.
It turns out that in the jargon of enzyme kinetics where the Michaelis-Menten
formula is widely used, half-saturation constants are seen as inverse
measures of the \emph{affinity} of an enzyme for its substrate. If
the analogy were that of flower or fruits being substrates, and pollinators
or frugivores being enzymes (i.e. facilitators), then $1/K=\beta/\omega$
would be the relative affinity of the animal for the flowers or fruits
of the plant. Now, if we decide instead to divide the numerator and
the denominator of the plant's numerical response by $\omega$, it
can be written like Holling's disc equation:

\begin{equation}
\frac{aN_{2}}{1+ahN_{2}}=\frac{\left(\nicefrac{\alpha\beta}{\omega}\right)N_{2}}{1+\left(\nicefrac{\beta}{\omega}\right)N_{2}}\label{eq:holling}
\end{equation}

\noindent where the rate at which the provider acquires benefits $a=\alpha\beta/\omega$,
is proportional to fruit or flower production $\alpha$, and to the
ratio at which they are used instead of wasted $(\beta/\omega)$,
e.g. the efficiency or affinity of the pollination or seed dispersal
process. The ``handling time'' of the plant becomes $h=1/\alpha$,
the average time it takes to create new flowers or fruits.

\begin{figure}
\begin{centering}
\includegraphics{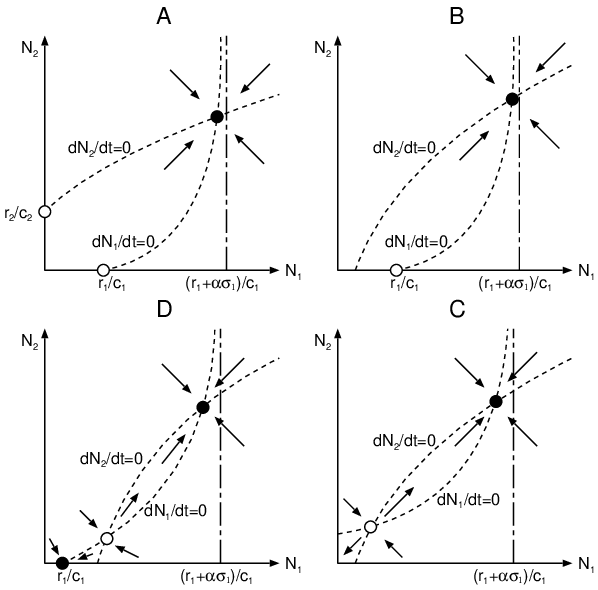}
\par\end{centering}

\caption{\label{fig:phaseplane2}Nullclines and equilibrium points in mutualisms
with exchange of resources for services, assuming linear self-limitation
for each species. The explanation is the same as in Figure \ref{fig:phaseplane1},
only this time the nullcline of the resource provider (species 1)
has an asymptote on its axis.}

\end{figure}

Using equation (\ref{eq:logistic}) to model the growth rates in the
absence of the mutualism in model (\ref{eq:unidir}), it is straightforward
to conclude that species 1 and 2 will respectively grow $(dN_{i}/dt>0)$
only if:

\begin{eqnarray}
N_{1} & < & \frac{r_{1}}{c_{1}}+\frac{\sigma_{o}\beta_{o}\alpha N_{o}+\sigma_{1}\beta\alpha N_{2}}{c_{1}(\omega+\beta_{o}N_{o}+\beta N_{2})}\label{eq:hypernull}\\
N_{1} & > & \frac{(c_{2}N_{2}-r_{2})(\omega+\beta_{o}N_{o}+\beta N_{2})}{\sigma_{2}\beta\alpha}\label{eq:paranull}
\end{eqnarray}

\noindent and decrease if the signs of the inequalities are respectively
reversed. The nullcline of the animal (ineq. \ref{eq:paranull} with
``='' instead of ``>'') is a parabola with the same properties
of the nullclines in the previous model (\ref{eq:nullcline}), only
that $\omega$ becomes $\omega+\beta_{o}N_{o}$. The plant's nullcline
(ineq. \ref{eq:hypernull} with ``='' instead of ``<'') differs
from the previous model as its graph is a rectangular hyperbola instead
of a parabola. The plant's nullcline has a single root on the plant
axis, $\frac{r_{1}}{c_{1}}+\frac{\sigma_{o}\beta_{o}\alpha N_{o}}{c_{1}(\omega+\beta_{o}N_{o})}$.
If this root is negative, the plant is an obligate mutualist of species
2 because its intrinsic growth rate is negative $(r_{1}<0)$, and
other means of pollination/seed dispersal (i.e. $\beta_{o}N_{o}>0$)
are insufficient to compensate the losses. On the other hand if this
root is positive, it may still be that the plant's intrinsic growth
rate is negative, yet pollination/seed dispersal not involving species
2 is enough to sustain the plant's population. The maximum abundance
that the plant could attain thanks to species 2 is limited by the
plant's nullcline asymptote at $N_{1}=(r_{1}+\sigma_{1}\alpha)/c_{1}$.
This means that if the plant's intrinsic growth rate is negative,
the rate of flower/fruit production $(\alpha)$ times the returns
$(\sigma_{1})$ from the mutualism, must overcome mortality $(\alpha\sigma_{1}>-r_{1})$,
otherwise the abundance of species 2 will not prevent the extinction
of the plant. Figure \ref{fig:phaseplane2} shows the graphs of the
nullclines.

Bearing in mind these details about the shapes of the nullclines,
the dynamics of the mutualism between the plant providing the food
and the animal providing the service are qualitatively the same as
in the case of both species providing material aid (e.g. food or nutrients),
as seen by comparing Figure \ref{fig:phaseplane2} with Figure \ref{fig:phaseplane1}.
The outcomes range from globally stable coexistence when both species
are facultative mutualists, to locally stable coexistence or no coexistence
when one or both are obligate mutualists.

\section*{Discussion}

Using a separation of time scales and the assumption of fast resource
dynamics, it was possible to derive simple models for mutualistic
interactions. In these models, the effect of one species abundance
on the growth rather of another, is grounded on microscopic principles,
rather than phenomenology. These \emph{numerical responses} display
decrease due to intra-specific competition, or because of diminishing
returns in the acquisition of benefits, enhancing the stability of
the interaction. My insistence of avoiding the term \emph{functional
response} and is entirely intentional. In the context of predation,
functional response is meant to refer to the per-capita predation
rate, whereas the numerical response describes the effect of prey
density on predator growth rates \citep{solomon-jae49,holling-aren61}.
While it is still correct to refer to functional responses with regard
to the consumption of resources provided by a mutualist (e.g. $\beta_{i}F_{j}$
in equation \ref{eq:food-food} is a type I functional response),
the models derived (e.g. equation \ref{eq:bidir}) describe the effect
of population densities on growth rates, not consumption rates. Thus,
the term numerical response is more appropriate in the present context.

In the case where each species produces a resource for the other,
intra-specific competition results from the fact that resource production
occurs at a finite rate per individual provider, independently of
its population size. If the population of the provider is kept constant,
this results in constant amount of resources provided per unit time,
that will be partitioned among the members of the other species, in
their role as consumers. If the consumer population is low, then each
individual receives a constant share, since the decay rate of the
resources is much larger than the rate at which they are consumed
$(\omega_{i}\gg\beta_{j}N_{j})$. This is no longer true when consumer
populations are large, which is when competition causes every individual
to get a share that decreases with the number of con-specifics \citep{schoener-tpb78}.

In the case where only one species provides the resource, that resource
is an \emph{organ} (e.g. flower, fruit) used by the provider (e.g.
plant) to \emph{extract} a service (e.g. pollination, seed dispersal)
from the consumer (animal). These organs must be regularly replaced
as they are used or decay, but like in the first scenario this happens
at a finite rate per individual no mater how large is its population.
If the population of the provider is kept constant, and the population
of the consumer is low, the rate at which the provider acquires benefits
per unit of consumer depends on the production to decay ratio $(\alpha_{i}/\omega_{i})$,
such that doubling the number of consumers doubles the benefits for
the plants. When the consumer population is large, providers cannot
regenerate the resource providing organs faster than the rate at which
they are used. For this reason, the more the provider helps the consumer
to grow, the lower its capacity to benefit from that increase, which
explains the diminishing returns.

\citet{fishman_hadany-tpb10} also used time scale assumptions in
order to model functional responses. However, their model is very
specific for pollination by bees, and they go into higher levels of
detail, by considering flower and patch states, and flower--nest traveling
times. Under certain assumptions, their result derives into the Beddington-DeAngelis
function, thus providing a mechanistic underpinning to the functional
forms used by \citet{holland_deangelis-ecology10}. While lacking
in detail, the models developed herein are more general, they can
encompass plant--frugivore mutualisms in addition to plant--pollinator
ones, when the resources are traded for services, or they can model
plant--mycorrhizae systems where organic compounds are traded for
essential nutrients.

I assumed that the consumption of resources enabling the mutualism
follow simple mass action laws. In reality, consumption is very likely
to display saturating functional responses (here the use of \emph{functional}
rather than \emph{numerical} is correct). In an interaction such as
frugivory, saturation could follow the disc equation mechanism \citep{holling-canent59b},
where the searching time of the consumer decreases with the number
of fruits, leading to an hyperbolic function of the number of flowers.
In pollination however, the fraction of time during which a flower
is not visited, i.e. the ``flower waiting time'', would decrease
with the number of pollinators which increase the ``flower working
time''. Thus in contrast with frugivory, pollination must consider
simultaneous saturation in plants and animals, and the Beddington-DeAngelis
function would be a reasonable choice describing flower use. Replacing
mass action laws with highly non-linear responses in the resource
dynamics will make it very difficult to derive simple results as those
presented. The absence of these complexities in the present formulation
does not however, belittles the approach taken, which stresses the
importance of considering the ephemeral nature of many kinds of resources
shared in mutualistic interactions. The fact that these resources
must be continuously regenerated at rates that are limited at the
individual level, causes dynamical bottlenecks in the acquisition
of benefits that ought to be considered, independently of the resource
consumption patterns.

The numerical responses are by itself not sufficient to stop population
growth at higher densities, they merely slow down population growth.
To see this, suppose that $G_{i}=-r$ i.e. first order mortality rates.
Substituting this in models (\ref{eq:bidir},\ref{eq:unidir}), the
population dynamics of a species can be written (in a compact form)
as:

\[
\frac{dN_{i}}{dt}=\left\{ \frac{aN_{j}}{b+N_{i}}-r\quad\textrm{or}\quad\frac{c+aN_{j}}{b+N_{j}}-r\right\} N_{i}
\]

\noindent where $a,b,c$ is everything that is not a variable. Suppose
that thanks to mutualism, the populations become very large ($N_{i}\gg b$
and $N_{j}\gg c/a$). In the first alternative populations will follow
$dN_{i}/dt\approx aN_{j}-rN_{i}$ and in the second case $dN_{i}/dt\approx(a-r)N_{i}$
($a>r$, otherwise the population would go extinct, contradicting
that mutualism caused it to be large). Both are linear systems characterized
by long term exponential growth, thus an equilibrium is not possible.
There are several ways to achieve population decrease at such high
densities. One is that mortality increases at higher than linear rates,
for example quadratically$-rN_{i}^{2}$, which by using (\ref{eq:logistic}).
Another possibility suggested by \citet{johnson_amarasekare-jtb13}
is to consider functions like above, but with population densities
raised to the second power in the denominator, simulating interference
among consumers. Interestingly, interference of the producer by the
consumer can enhance stability too. I assumed that resource consumption
does not cause any damage to the provider, but now suppose that resource
provision can be inhibited by too many consumers. A simple way to
frame this is that the rate at which producer $j$ makes resources
in equation (\ref{eq:food-food}) is $\alpha_{j}N_{j}/(1+\gamma_{i}N_{i})$,
where $\gamma_{i}$ scales the inhibitory effect of consumers. Using
the steady-state assumption, the resource levels will be:

\[
F_{j}\approx\frac{\alpha_{j}N_{j}}{(1+\gamma_{i}N_{i})(\omega_{j}+\beta_{i}N_{i})}
\]

\noindent where the denominator is quadratic in the consumer population.
Thus, the rate of mutualism in equation (\ref{eq:res-res}) can be
lower than first order mortality rates when the populations are large,
like in \citet{johnson_amarasekare-jtb13} model. This scheme also
applies to the resources for services model \emph{mutatis mutandis}.

Provided that conditions of fast resource dynamics and slow population
dynamics apply, models (\ref{eq:bidir},\ref{eq:unidir}) can be generalized
to larger communities. Adding as many consumption terms as consumers
in equations (\ref{eq:food-food},\ref{eq:floresfrutas}), the steady
state level of resources given by the provider will be:
\begin{equation}
F_{j}\approx\frac{\alpha_{j}N_{j}}{\omega_{j}+\sum_{i}\beta_{ji}N_{i}}\label{eq:multispecies}
\end{equation}

\noindent so, a consumer that eats several of these resources (e.g.
pollinators) will follow the dynamics:

\[
\frac{dN_{i}}{dt}=\left\{ G_{i}(\cdot)+\sum_{j}\frac{\sigma_{ji}\beta_{ji}\alpha_{j}N_{j}}{\omega_{j}+\sum_{k}\beta_{jk}N_{k}}\right\} N_{i}
\]

\noindent which is a multi-resource extension of \citet{schoener-tpb78}
competition models. For a species that gives resources but receives
services instead (e.g. flowering plants), the dynamics will follow:

\[
\frac{dN_{j}}{dt}=\left\{ G_{j}(\cdot)+\frac{\alpha_{j}\sum_{j}\sigma_{ji}\beta_{ji}N_{i}}{\omega_{j}+\sum_{i}\beta_{ji}N_{i}}\right\} N_{j}
\]

\noindent where several benefits are pooled together in a multi-species
numerical response.

The parameters of these numerical responses are rates which are very
likely to be related by means of trade-offs. Consider for example,
fragile flowers or fruits (high $\omega_{i}$) that are cheaper to
produce (large $\alpha_{i}$); or that there is an inverse relation
between the consumption rate $(\beta_{i})$ and assimilation ratio
$(\sigma_{i}$), i.e. an efficiency trade-off. In multi-specific contexts,
the finite nature of the individuals constrain their consumption rates
$(\beta_{ji})$, such that the increase of one causes the decrease
in others. Some of these parameters, like the provision of benefits,
can be related with the intrinsic growth rates $G_{i}$ \citep{johnson_amarasekare-jtb13};
for example $dG_{i}/d\alpha_{i}<0$, because the energy used to make
resources for other species could have been used to improve the provider's
capacity to live on its own. Since the production of benefits by species
$i$ affects the dynamics of species $j$ (\ref{eq:bidir},\ref{eq:unidir}),
the coevolution of mutualism can be investigated. Furthermore, if
the costs of providing benefits can go as far as changing the sign
of $G_{i}$ from positive to negative, the mechanistic framework described
by this paper can be used to study the evolutionary origin of obligate
mutualisms.

The use of time scale arguments is widespread in the theoretical biology.
In ecology, the derivation of the competitive Lotka-Volterra equations
by \citet{macarthur-tpb70} is a well known example. A lesser known
example (but more relevant here), are the competitive models derived
by \citet{schoener-tpb78}, which consider the partition of resource
inflows. The derivation of the disc equation \citep{holling-canent59b}
requires the fractioning of a predation cycle, a period of time that
is embedded into the much longer time scale of the population dynamics.
The scenarios suggested herein are far from exhaustive. One goal of
this note is to see, to what extent, simple time scale assumptions
can help unify consumer-resource, mutualism and competition theories.
Another goal concerns the mechanistic derivation of generic models,
with few complexities, but based on parameters that can be potentially
measured such as rates of flowering or nectar production and decay,
and consumption rates.

\section*{Acknowledgements}

I thank Luis Fernando Chaves, Bart Haegemann, Michel Loreau, Claire
de Mazancourt, Jarad Mellard, Shaopeng Wang, Francisco Encinas-Viso
for their helpful comments and suggestions. This work was possible
thanks to the support of the TULIP Laboratory of Excellence (ANR-10-LABX-41).

\bibliographystyle{chicago}
\bibliography{timescales}

\end{document}